# Overview

## Title

Simple RGC: ImageJ plugins for counting retinal ganglion cells and determining the transduction efficiency of viral vectors in retinal wholemounts

## Authors

Tiger Cross[1,†], Rasika Navarange[1,†], Joon-Ho Son[1,†], William Burr[1,†], Arjun Singh[1,†], Kelvin Zhang[1,†], Miruna Rusu[1], Konstantinos Gkoutzis[1], Andrew Osborne[2,*] and Bart Nieuwenhuis[2,3,*]

[†] These authors have contributed equally to this work.
* These authors share senior authorship and correspondence (Andrew Osborne – E-mail: ao361@cam.ac.uk, Tel.: +44 1223 334027; Bart Nieuwenhuis – E-mail: bn246@cam.ac.uk, Tel.: +44 1223 334027).

## Affiliations

[1] *Department of Computing, Imperial College London, 180 Queen's Gate, South Kensington, London, SW7 2AZ, UK*

[2] *John van Geest Centre for Brain Repair, Department of Clinical Neurosciences, University of Cambridge, Forvie Site, Robinson Way, Cambridge, CB2 0PY, UK*

[3] *Laboratory for Regeneration of Sensorimotor Systems, Netherlands Institute for Neuroscience, Royal Netherlands Academy of Arts and Sciences (KNAW), Meibergdreef 47, 1105 BA, Amsterdam, the Netherlands*

**Abstract**

'Simple RGC' consists of a collection of ImageJ plugins to assist researchers investigating retinal ganglion cell (RGC) injury models in addition to helping assess the effectiveness of treatments. The first plugin named 'RGC Counter' accurately calculates the total number of RGCs from retinal wholemount images. The second plugin named 'RGC Transduction' measures the co-localisation between two channels making it possible to determine the transduction efficiencies of viral vectors and transgene expression levels. The third plugin named 'RGC Batch' is a batch image processor to deliver fast analysis of large groups of microscope images. These ImageJ plugins make analysis of RGCs in retinal wholemounts quick, consistent, and less prone to unconscious bias by the investigator. The plugins are freely available from the ImageJ update site https://sites.imagej.net/Sonjoonho/.

**Introduction**

The retina is a thin layer of tissue located at the back of the eye. Its purpose is to receive light focused from the lens, and to convert the light into neural signals to be sent onto the brain (**Figure 1A**). The cells that transport these neural signals are known as retinal ganglion cells (RGCs). Injuries to RGCs occur in a variety of ocular disorders that can lead to blindness, and there is a large research community investigating ways to protect these cells from damage.

Retinal wholemounts, the intact removal of the retina from the eye, are the favoured technique to quantify RGC number and to assess the impact of protective treatments, such as gene-therapies (**Figure 1B**). However, these studies require the processing of hundreds of images, each containing hundreds of cells (**Figure 1C**). Manual quantification is labour-intensive, time-consuming and prone to user variation. It is clear that automated quantification improves productivity, accuracy, and reproducibility in research.

There are expensive image processing software that can count and measure co-localisation (e.g. Zen, LAX, Volocity, Imaris). Open-source computational tools for automated image analysis are also becoming increasingly available. For instance, several software packages that count the number of RGCs in naive and injured tissues have been developed in the last five years [1–4]. There are also freely available co-localisation programs [5–9], but none are optimised for determining viral transduction efficiencies of RGCs in retinal wholemounts. Furthermore, only a few of the above mentioned programmes have the option to batch-process a large number of images.

ImageJ [10–13] is an open-source program that is commonly used for image analysis in biomedical sciences. ImageJ has a variety of built-in tools to aid image analysis and enables users to create and run their own plugins. However, many ImageJ plugins still require a level of manual analysis with a repetitive workflow. We have therefore developed three ImageJ plugins to automate image analysis from retinal wholemount images.

(1) RGC Counter: this plugin detects and outlines individual RGCs, providing a quick and accurate method to quantify the total number of cells from a single-colour image.

(2) RGC Transduction: this plugin calculates the co-localisation between two colour channels and generates the following output per image: (a) total number of RGCs; (b) number of transduced RGCs (co-localised RGCs); (c) viral transduction efficiency (percentage co-localisation); (d) fluorescence intensity measurement of the transgene for each transduced RGC. In addition to being able to specify cell size and reduce noise, this plugin makes use of an optional feature that subtracts transduced axons to avoid false positives.

(3) RGC Batch: this plugin is a batch image processor compatible with both 'RGC Counter' and 'RGC Transduction' to improve workflow.

The automated analysis of the first two ImageJ plugins can be previewed in real-time to assist in establishing the parameters. The plugin results can be viewed in ImageJ or exported as CSV and XLSX files.

Taken together, we have created a set of precise, freely available and user-friendly ImageJ plugins for the retinal research community. These plugins will be particularly useful for researchers investigating neuroprotective strategies or assessing the suitability of their gene therapy vectors for inner retina transduction.

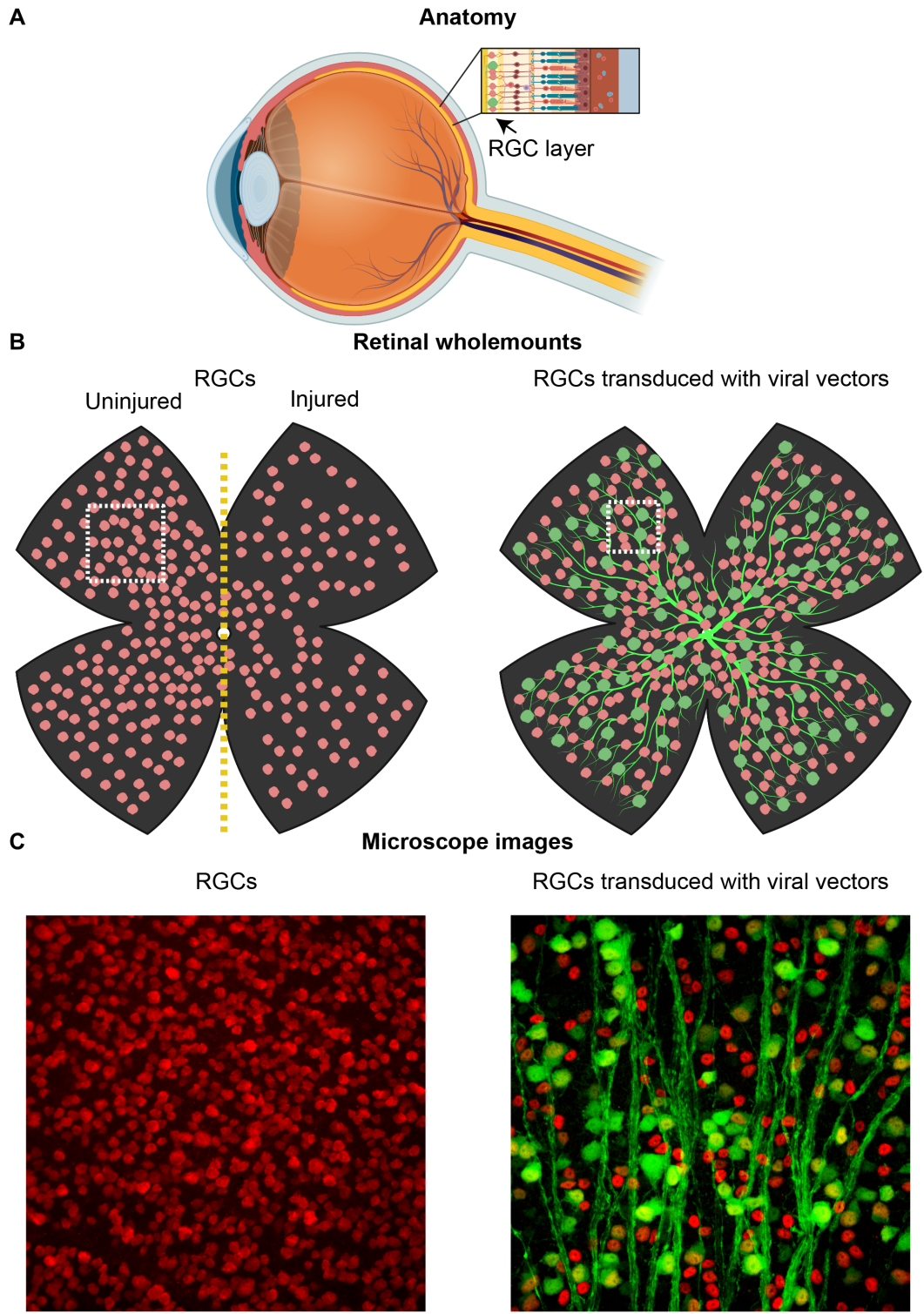

**Figure 1:** A schematic representation of the eye and retina to illustrate where the images required for the plugin originate. (A) The retina is located at the back of the eye and contains multiple cell types. The retinal ganglion cells (RGCs - red) are responsible for communicating information from the eye to the brain via their long axons. (B) A set of retinal wholemounts illustrating individual RGCs (red) or transduced RGCs (green on top of red). Note the viral vector also labels the RGC axons. (C) Fluorescence microscopy images showing RGCs labelled with RBPMS (red – left) or Brn3A (red – right) in addition to fluorescence generated from the viral vector (green – right). White boxes indicate magnified regions for analysis. Retinal transduction was achieved using a rAAV2-hSYN-eGFP viral vector. Images were taken on a Leica DMi8 microscope with a 20x objective (left) or Leica SPE confocal microscope at 40x (right). Schematics were created with BioRender.com.

## Implementation and architecture

### Overview

ImageJ supports a wide variety of languages for its application-programming interface (API), including Java and Python (Jython). These languages are used to produce a JAR plugin file for users to load into ImageJ. The 'Simple RGC' plugins are written in Kotlin [14], a statically typed programming language designed to fully inter-operate with Java [15]. With ImageJ being developed in Java, its Java API documentation is much more thorough compared to its other supported languages. Kotlin allowed us to leverage this whilst gaining concise syntax and the addition of powerful functional programming constructs.

The code for 'Simple RGC' is split into an abstraction provided by the ImageJ API: commands and services. Commands are designed to be used for one-off operations, while services are intended to be re-used across images. As a result, the code to handle the user interfaces (UIs) of 'Simple RGC' is provided by commands and the image segmentation steps are encapsulated within services. This structure reduced the amount of duplicated code in our plugins and increases its maintainability.

### Installation

'Simple RGC' requires the user to download ImageJ prior to installation. ImageJ, and associated-bundles with pre-installed plugins such as Fiji, can be downloaded at:

https://imagej.net/Fiji/Downloads. It is recommended that current ImageJ users upgrade to the latest software version by selecting 'Help > Update ImageJ...' from the ImageJ menu. The ImageJ plugins that are part of 'Simple RGC' become accessible by adding our ImageJ update site http://sites.imagej.net/Sonjoonho/ to the install of ImageJ. A detailed description on how to follow an update site can be found on the ImageJ wiki page: https://imagej.net/Following_an_update_site. In brief, select 'Help > Update... > Manage update sites > Add update site'. Next, insert 'http://sites.imagej.net/Sonjoonho/' into the URL box and tick the corresponding box. Afterwards, click 'Update URLs > Apply changes > OK' and restart ImageJ. The 'Simple RGC' plugins are now installed and will be accessible from the 'Plugins' menu in ImageJ.

### *User interface and compatibility with microscopy images*

We chose to use the ImageJ graphical user interface and API for 'RGC Counter' and 'RGC Transduction', as opposed to implementing a new UI. This was done to ensure that we developed a UI with a similar style and convention to other ImageJ plugins, reducing the learning curve for ImageJ users. This allows for a straightforward and friendly user experience.

All plugins are designed to be compatible with TIFF, PNG and proprietary formats such as Leica Image Files (LIFs). This allows direct export of native images from a microscope to the plugin.

### *RGC Counter*

'RGC Counter' quantifies the total number of cells in an image. The identified cells become highlighted in ImageJ and the cell counts are presented.

Following ImageJ convention, 'RGC Counter' operates on an image that has already been opened. To run the plugin, the user should select 'Plugins > Simple RGC > RGC

Counter' from the ImageJ menu. The user then has the option to manually tune parameters based on cell size and image quality (**Figure 2 – User Interface**). It is recommended to use the preview button during the initial setup to optimise parameters for the most accurate detection of cells. The plugin will remember the most recently used parameters to improve workflow.

After the user clicks "OK" on the UI, the plugin begins to run. 'RGC Counter' runs a series of steps in order to successfully segment and identify each cell in the selected morphology channel of a fluorescence microscopy image. The steps of the image processing are shown in **Figure 2 – Image processing** and they are as follows:

1. The colour channel selected by the user is extracted and a grayscale filter is applied.
2. A local threshold is applied to the grayscale image. At a high level, this operates by moving a small window over the image and setting pixels to white if they are above a certain threshold, and black if they are below. The Niblack's algorithm [16] is implemented at the local thresholding step, as we found that it outperformed Bernsen's [17] and Otsu's [18] on accuracy and image processing speed (data not shown).
3. The third step is optional, and when selected the plugin attempts to remove axons by (a) identifying them using the 'Ridge Detection' plugin [19], and (b) masking them out. It is not necessary to select this optional step when the loaded image contains relatively round structures (e.g. Brn3a, RBPMS, NeuN, DAPI and Hoechst stains), but could improve the quantification output when both cells and neurites are visualised (e.g. eGFP or Tuj1-positive cells). Please note that activation of 'Exclude axons' increases the time duration for image processing.
4. A Gaussian blur [20] is applied to the image in order to merge erroneously separated cells.
5. A global threshold is applied to clarify the blurred image. Otsu's algorithm [18] was used at the global thresholding step. We found that Otsu's performed similar to Moment's [21] and Shanbhag's [22] in sharpening images.

6. The watershed algorithm [23] is applied to separate connected cells into individual entities.

7. The final step is based on particle analysis to filter out structures (e.g. debris or artefacts of staining) that are not within the cell diameter range specified by the user, and importantly the identified cells are highlighted and quantified.

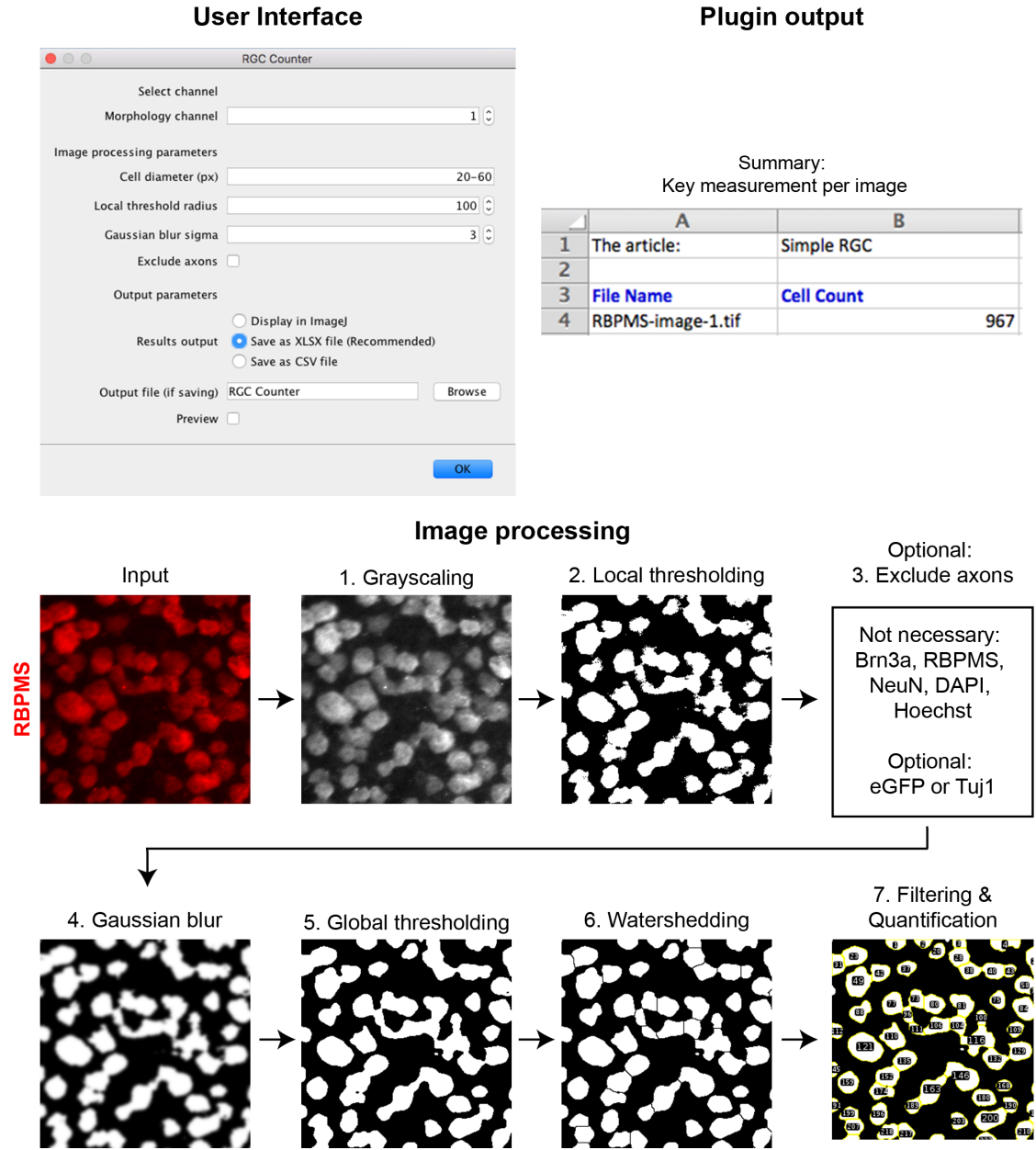

**Figure 2:** Architecture of ImageJ plugin 'RGC Counter'. The user interface, the output in XLSX format, and the image processing steps for this plugin are shown.

The results output of 'RGC Counter' can be view in ImageJ's ROI Manager. The results output can also be directly saved as XLSX or CSV file. These saved output files capture the total cell count (**Figure 2 – Plugin output**) as well as the user-set parameters for enhanced documentation.

**Note 1:** The cell diameter range in pixels can be manually determined in two simple steps: 1) Remove the scale from the open image by selecting 'Analyze > Set Scale… > Click to Remove Scale > OK' in the ImageJ menu; 2) Determine the cell diameter range by drawing a straight line on top of individual cells of interest by selecting 'Straight Line Selection Tool' in the ImageJ toolbar and afterwards measure by clicking 'Analyze > Measure'. Alternatively, the user can estimate the cell diameter range in pixels by opening the image and using the 'preview' button from the UI of the plugin.

### *RGC Transduction*

'RGC Transduction' identifies the co-localisation between two colour channels with the rationale to identify cells that have been transduced by a viral vector. Moreover, fluorescence intensity measurements are recorded in *all* channels of an open image. The cells classified as co-localised are highlighted on the image, and the results are tabulated.

'RGC Transduction' requires that an image is opened in ImageJ in order to run the plugin. To run the plugin, the user should select 'Plugins > Simple RGC > RGC Transduction' from the ImageJ menu. The user is required to select the channels for co-localisation analysis, and to specify the image-processing parameters (**Figure 3 – User Interface**). It is recommended to use the preview button during the initial setup to optimise parameters for the most accurate measurements of transduced cells. The plugin will remember the most recently used parameters to improve workflow.

'RGC transduction' runs a series of steps that are highlighted in **Figure 3 – Image processing.** Initially, the image is split into two separate single-channel images, corresponding to the channels for morphology and transduction selected by the user. With the need to identify individual cells across the morphology and transduced channels, the same pre-processing and segmentation pipeline as 'RGC Counter' is applied separately to both channels. Afterwards, co-localisation between the two channels is calculated. A cell is classified as transduced if at least 50% of the pixels in the morphology channel are also covered by pixels in the transduction channel. This threshold of colocalisation achieved the best performance on the investigated images containing RGCs and the transgene eGFP.

The results output of 'RGC Transduction' can be view in ImageJ's ROI Manager or can be directly saved as XLSX or CSV files. We recommended saving the results in the XLSX format because it will generate one organised file with multiple tables, in contrast to multiple CSV files. Highlights of measurements performed by 'RGC Transduction' are shown in **Figure 3 – Plugin output.** The tab called 'summary' will show key measurements for the image, including 'total number of cells', 'number of transduced cells', 'transduction efficiency (%)', 'average morphology area', and various fluorescence intensities. The tab entitled 'transduced cell analysis' will show all the metrics for each individually transduced cell. The tab named 'Parameters' contains the image-processing parameters that were selected for image analysis.

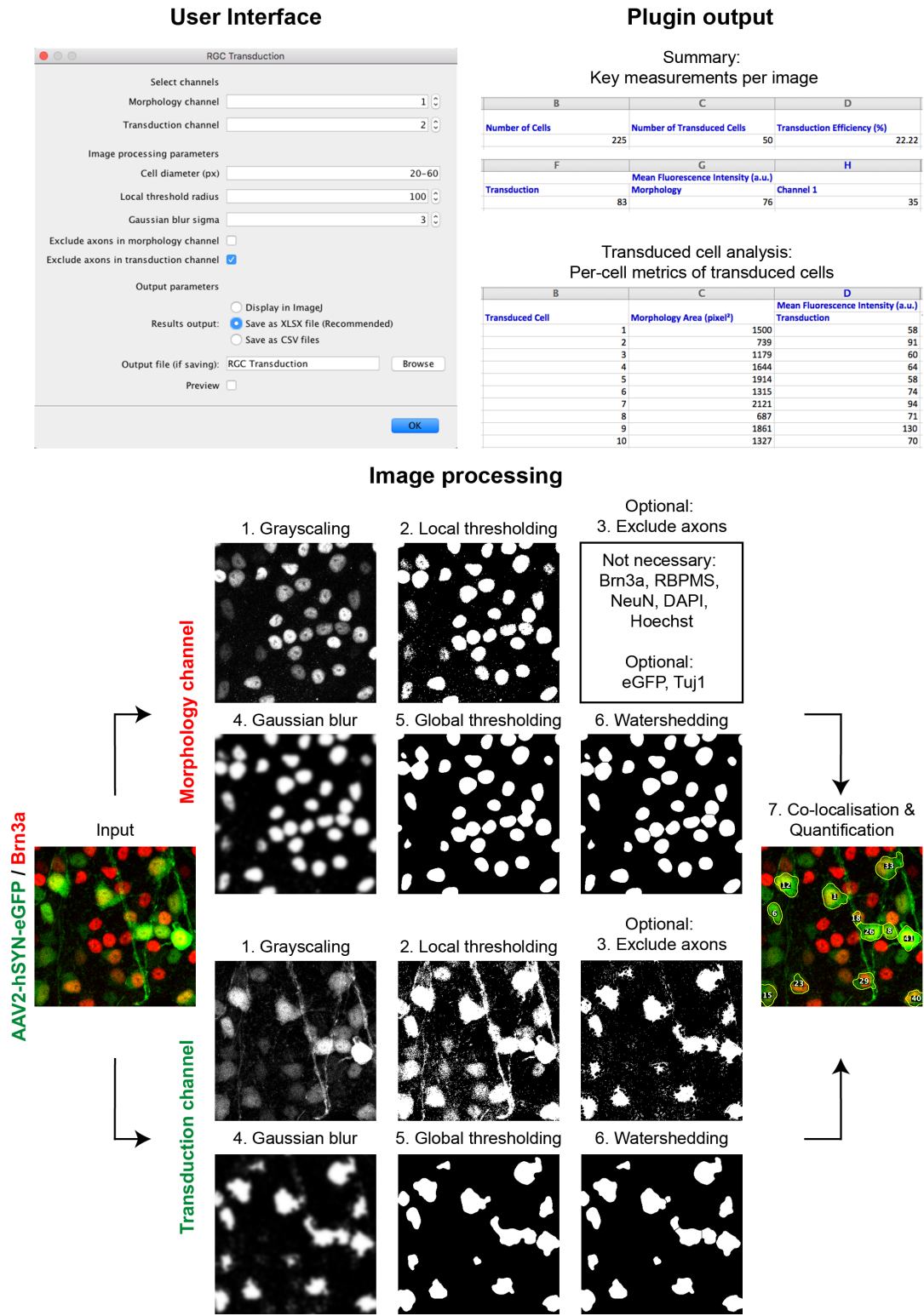

**Figure 3:** Architecture of ImageJ plugin 'RGC Transduction'. The user interface, the key parts of the output in XLSX format, and the image processing steps for this plugin are shown.

### *RGC Batch*

'RGC Batch' is a batch-image processor that allows the user to run multiple image files through 'RGC Counter' or 'RGC Transduction' with minimal user interaction or repetition.

The UI is not built using ImageJ's command interface as done for 'RGC Counter' and 'RGC Transduction'. Instead, we built our own custom UI using the Java Swing framework to circumvent ImageJ UI restrictions. This allowed us to make separate tables for both 'RGC Counter' and 'RGC Transduction'. The user interface of RGC Batch is shown in **Figure 4 – User Interface.**

From a technical perspective, 'RGC Batch' simply acts as a wrapper to both 'RGC Counter' and 'RGC Transduction' making it capable of running a collection of images from a selected folder as illustrated in **Figure 4 – Workflow**.

The results generated by 'RGC Batch' can be saved in one XLSX file (recommended) or multiple CSV files. Highlights of measurements performed by 'RGC Batch' are shown in XLSX format in **Figure 4 – Plugin output.** The output of 'RGC Batch' when running 'RGC Counter' is largely similar to that of 'RGC Counter', only adding additional rows for each image processed. The output of 'RGC Batch' when running 'RGC Transduction' has notable differences to that of 'RGC Transduction'. The main reason for this is to ensure that the metrics for each cell are clearly organised per individual image. The output of 'RGC Batch' when running 'RGC Transduction' consists of tables for each measurement and the individual metrics are shown for each cell per image.

**Note 2:** Before running 'RGC Batch', ensure that all images have the same channel ordering and that microscope settings are consistent across all images.

**Note 3:** The generation of an output file is more time consuming when one of the following image processing parameters are selected: 'Exclude axons', 'Exclude axons in morphology channel', 'Exclude axons in transduction channel'.

## User Interface

**RGC Batch - RGC Counter**

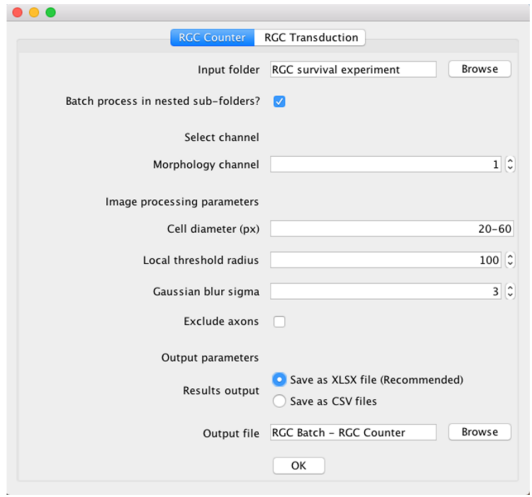

**RGC Batch - RGC Transduction**

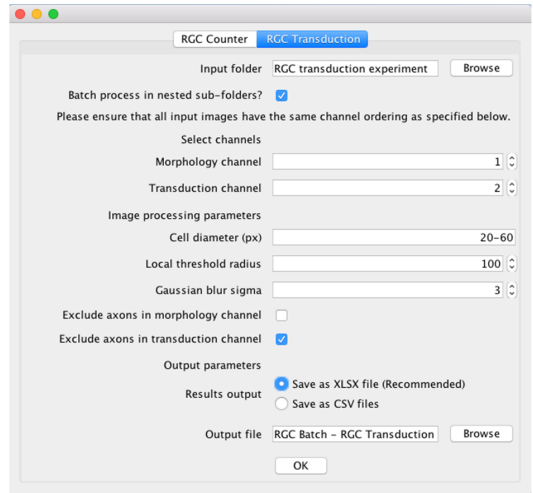

## Workflow

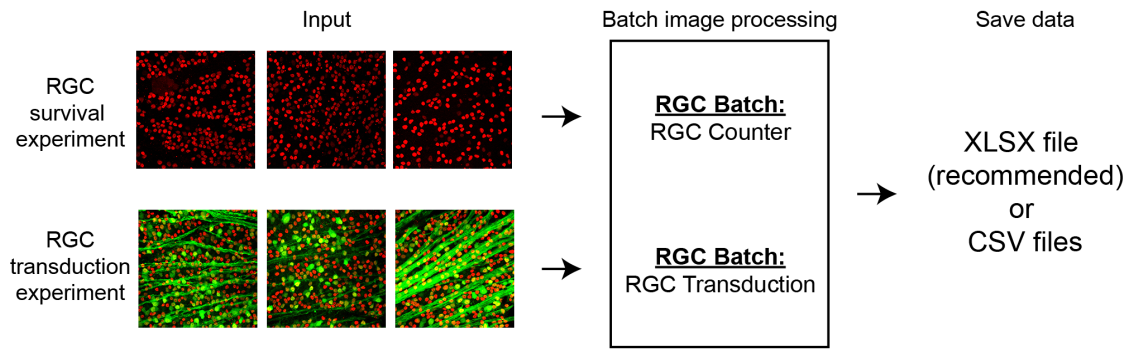

## Plugin output

**RGC Batch - RGC Counter**

Summary:
Key measurements per image

| | A | B | C | D |
|---|---|---|---|---|
| | The article: | Simple RGC | | |
| | **File Name** | **Cell Count** | | |
| | RBPMS-image-1.tif | 583 | | |
| | RBPMS-image-2.tif | 622 | | |
| | RBPMS-image-3.tif | 655 | | |
| | RBPMS-image-4.tif | 649 | | |
| | RBPMS-image-5.tif | 578 | | |
| | RBPMS-image-6.tif | 678 | | |

**RGC Batch - RGC Transduction**

Summary:
Key measurements per image

| | A | B | C | D |
|---|---|---|---|---|
| | **File Name** | **Number of Cells** | **Number of Transduced Cells** | **Transduction Efficiency (%)** |
| | Transduced-image-1.tif | 225 | 59 | 26.22 |
| | Transduced-image-2.tif | 310 | 78 | 25.16 |
| | Transduced-image-3.tif | 315 | 53 | 16.83 |
| | Transduced-image-4.tif | 324 | 27 | 8.33 |
| | Transduced-image-5.tif | 256 | 45 | 17.58 |
| | Transduced-image-6.tif | 365 | 66 | 18.08 |
| | Transduced-image-7.tif | 200 | 81 | 40.5 |

Transduced cell analysis:
Mean fluorescence intensity for each transduced cell per image

| | A | B | C | D |
|---|---|---|---|---|
| | **Transduced Cell** | **Transduced-image-1** | **Transduced-image-2** | **Transduced-image-3** |
| | 1 | 59 | 146 | 62 |
| | 2 | 62 | 95 | 62 |
| | 3 | 68 | 74 | 59 |
| | 4 | 131 | 92 | 112 |
| | 5 | 127 | 125 | 124 |
| | 6 | 71 | 91 | 89 |
| | 7 | 49 | 90 | 78 |
| | 8 | 49 | 224 | 91 |

**Figure 4:** Architecture of ImageJ plugin 'RGC Batch'. The user interface of 'RGC Batch' allows the selection of either 'RGC Counter' or 'RGC Transduction'. The workflow is shown as a schematic diagram, and key outputs in XLSX format are shown.

## Quality control

The Simple RGC ImageJ plugins were validated by comparing the output of each plugin to manual quantification performed by six investigators. The time required to manually process each image was also measured.

### *Validation of RGC Counter*

RGCs from mouse retinal wholemounts were immunolabelled with the RGC marker RBPMS, and images were captured using an epifluorescent microscope at 20x magnification. The number of RGCs per image was manually counted in six images from uninjured- and injured retinal wholemounts by six independent investigators.

ImageJ plugin 'RGC Counter' detects and highlights RBPMS-positive cells from the image file and separates overlapping cells (**Figure 5A and C**). Pearson product-moment correlation analysis demonstrated a strong positive linear correlation between automated- and manual quantifications (**Figure 5B and D**). The Pearson correlation coefficient (r) was 0.961 for the quantifications of both uninjured- and injured retinal wholemounts images. The time required to manually quantify one image ranged from 2 to 16 minutes, depending on the investigator and number of RGCs per image (**Figure 5E and F**). In contrast, the automated quantification via 'RGC counter' took less than 20 seconds (**Figure 5E and F**). Taken together, the ImageJ plugin 'RGC Counter' quantifies the number of RGCs accurately and rapidly.

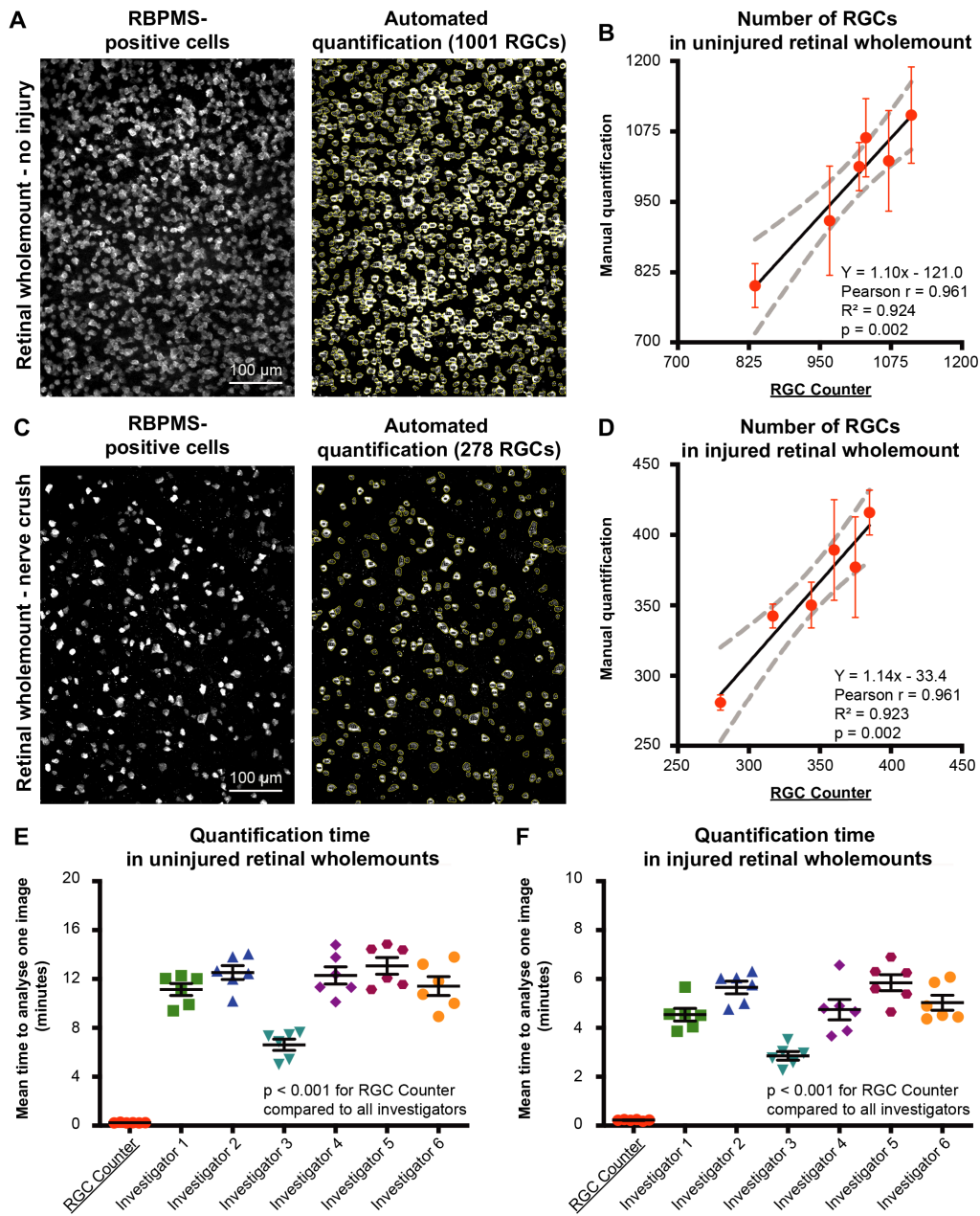

**Figure 5:** Assessment of RGC number by the ImageJ plugin 'RGC Counter'. RGCs in mouse retinal wholemounts were visualised using the immunohistochemical marker RBPMS. (A) Fluorescence microscopy image of an uninjured retinal wholemount (left) and automated quantification of RGCs by 'RGC Counter' (right). (B) Comparison between the number of RGCs detected by the 'RGC Counter' plugin and manual cell counts by six investigators from images of an uninjured retinal wholemount [Y = 1.10x − 121.0, Pearson's r = 0.961, $R^2$ = 0.924, p = 0.002, Pearson product-moment correlation]. (C) Fluorescence microscopy image of a retinal wholemount from an animal that underwent an optic nerve crush injury (left) and automated quantification of RGCs by 'RGC Counter' (right). (D) Comparison between the number of RGCs detected by the 'RGC Counter' plugin and manual cell

counts by six investigators from images of an injured retinal wholemount [Y = 1.14x – 33.4, Pearson's r = 0.961, $R^2$ = 0.923, p = 0.002, Pearson product-moment correlation]. (E – F) Time required to analyse a single image of the uninjured and injured retinal wholemounts, respectively [Uninjured: $F(6, 35)$ = 65.2, p < 0.001 for all investigators compared to RGC counter, ANOVA with Dunnett's multiple comparisons test; Injured: $F(6, 35)$ = 50.9, p < 0.001 for all investigators compared to RGC Counter, ANOVA with Dunnett's multiple comparisons test]. The Pearson product-moment correlation and ANOVA data are shown as average ± 95% CI and average ± SEM, respectively. Images were taken on a Leica DMi8 microscope with 20x objective.

### *Validation of RGC Transduction*

The retina had been transduced with a viral vector encoding enhanced green fluorescent protein (eGFP). RGCs from a mouse retinal wholemount were immunolabelled with the RGC marker Brn3A, and images were captured using a confocal microscope at 40x magnification. Twelve images (3 images per retinal quadrant) were captured to provide a global overview of the gene therapy effectiveness.

ImageJ plugin 'RGC Transduction' measures the co-localisation between the transduction (eGFP-positive cells) and morphology channel (Brn3a-positive cells) and then highlights and measures the number of transduced RGCs (**Figure 6A**). Pearson product-moment correlation analysis demonstrated a positive linear correlation between automated- and manual quantification (**Figure 6B and 3C**). The Pearson correlation coefficient (r) was 0.712 between the two quantification methods to determine the number of transduced RGCs (**Figure 6B**). Expectedly, the correlation is not as high as 'RGC Counter' due to the need to overlay two channels, and variability in transduction levels and morphology. Consistent with plugin 'RGC Counter', quantification of the total number of RGCs in the morphology channel was highly accurate and resulted in a Pearson correlation coefficient (r) of 0.995 (**Figure 6C**). The transduction efficiency of a viral vector is calculated by dividing the number of transduced RGCs by the total number RGCs – displayed in the output table of the ImageJ plugin. The automated quantification resulted in an average transduction efficiency of 40.2 ± 3.0 % for the viral vector across all images (**Figure 6D**). This outcome was not statistically significantly different from manual measurements calculated by the six investigators (**Figure 6D**). This result demonstrates that 'RGC Transduction' is sensitive enough to accurately determine the viral transduction efficiency of RGCs in retinal wholemounts.

As an added benefit, we also programmed the plugin to measure the transgene fluorescence intensity from each of the transduced RGCs (see **Figure 2 – Plugin output** and **Figure 4 - Plugin output**). This provides the user with a wider range of information, such as strength of transgene expression in addition to transduction percentage. This is a key advantage of using a plugin to enhance research output. When investigators were asked to determine the transgene fluorescence intensity of 15 transduced RGCs per image, cells chosen at random, the results varied greatly between the six investigators (**Supplementary figure 1**) also highlighting difficulties with reproducibility when completing image analysis manually.

Furthermore, the plugin vastly improved workflow speed. The automated quantifications via 'RGC Transduction' quantified the number of transduced RGCs, the total number of RGCs and the fluorescence intensity per cell within 25 seconds for each image. In contrast, it took investigators between 4 and 17 minutes per image to perform this task (**Figure 6E**).

Taken together, the ImageJ plugin 'RGC Transduction' reduces the amount of time needed for image analysis, accurately determines the number of transduced RGCs, the total RGC population, percentage viral transduction efficiency, and the individual transduction strength for each RGC.

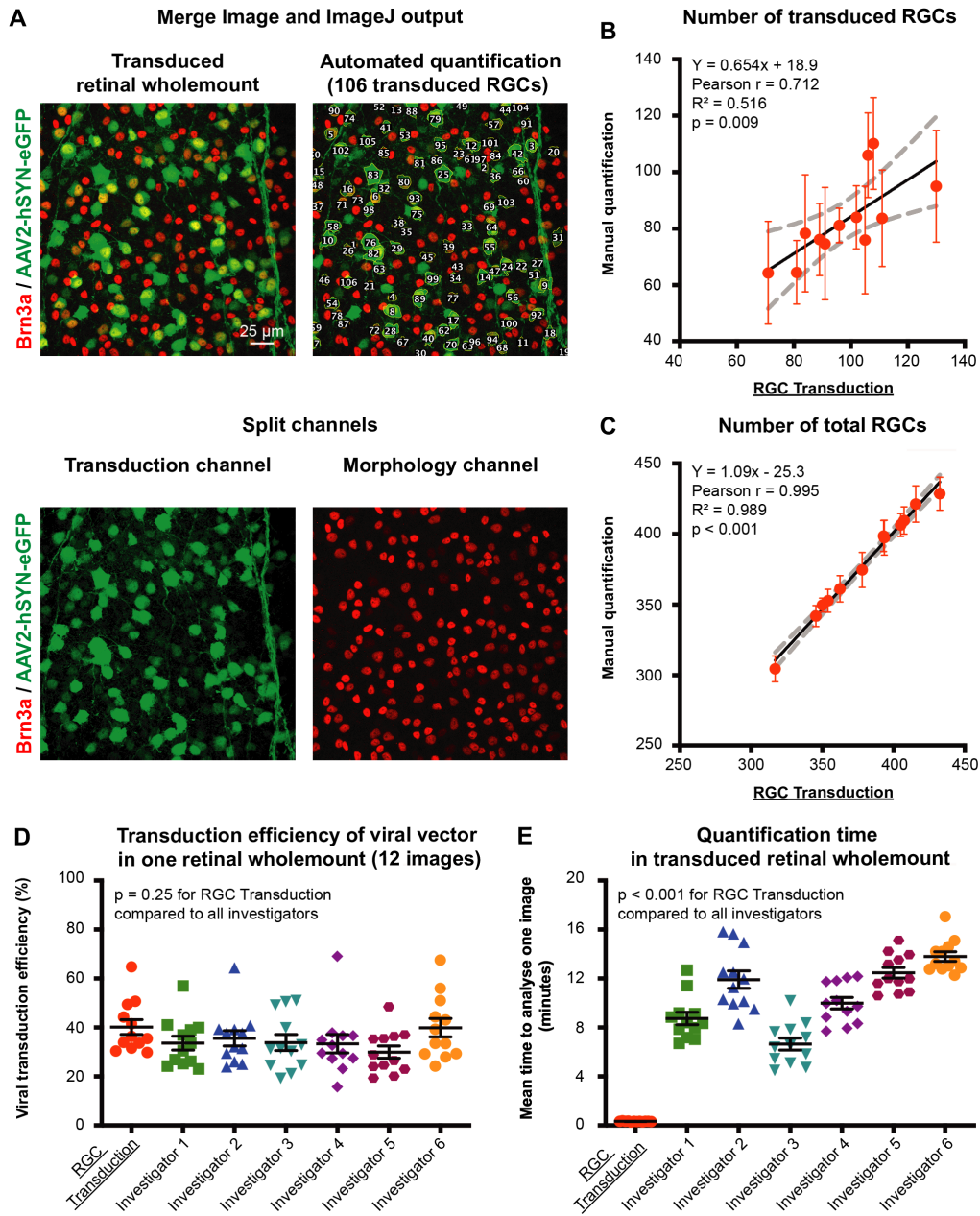

**Figure 6:** Assessment of the transduction efficiency of viral vectors for RGCs by the ImageJ plugin 'RGC Transduction'. The viral vector AAV2-hSYN-eGFP was delivered to the mouse retina via intravitreal injection (5x10E9 viral particles/eye). One month after injection, the retinal wholemount was prepared and RGCs were visualised using the immunohistochemical marker Brn3a. (A) Fluorescence microscopy image of a transduced retinal wholemount (top left) and automated quantification of transduced RGCs by plugin 'RGC Transduction' (top right). The merged image consisting of two colours is also shown as split channels for eGFP-positive cells in green (bottom left) and Brn3A-positive cells in red (bottom right). (B - C) Comparison between the results of the 'RGC Transduction' plugin and manual quantifications by six investigators for the number of transduced RGCs and total

number of RGCs, respectively, in 12 images from one transduced retinal wholemount [Number of transduced RGCs: Y = 0.65x + 18.9, Pearson's r = 0.718, $R^2$ = 0.516, p = 0.009, Pearson product-moment correlation; Number of total RGCs: Y = 1.09x − 25.3, Pearson's r = 0.995, $R^2$ = 0.989, p < 0.001, Pearson product-moment correlation]. (D) Transduction efficiency of the viral vector for RGCs in each analysed image [$F_{(6, 77)}$ = 1.33, p = 0.25 for all investigators compared to 'RGC Transduction', ANOVA with Dunnett's multiple comparisons test]. (E) Time required to analyse a single image of a transduced retinal wholemount [$F_{(6, 77)}$ = 92.9, p < 0.001 for all investigators compared to 'RGC Transduction', ANOVA with Dunnett's multiple comparisons test]. The Pearson product-moment correlation and ANOVA data are shown as average ± 95% CI and average ± SEM, respectively. Images were taken on a Leica SPE confocal microscope with 40x objective.

## *Validation of RGC Batch*

The functionality of ImageJ plugin 'RGC Batch' was validated by comparing the output of this batch image processor to the results of 'RGC Counter' and 'RGC Transduction'.

The results generated by 'RGC Batch' were identical to 'RGC Counter' (**Figure 7A**) and 'RGC Transduction' (**Figure 7B**). This demonstrates that 'RGC Batch' can process multiple images with the same high level of accuracy as the stand-alone plugins. This third plugin makes automated quantification of large data sets more user-friendly by avoiding a repetitive workflow when all images are taken using identical microscope settings.

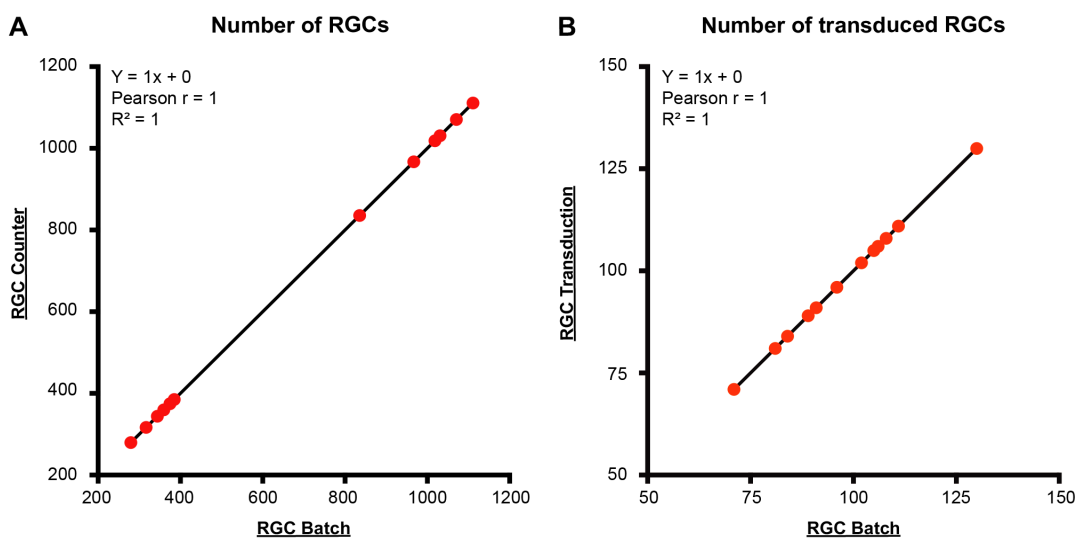

**Figure 7:** Validation of the ImageJ plugin 'RGC Batch'. Fluorescence microscopy images were analysed using the same user-set parameters and were either processed in one batch by 'RGC Batch' or assessed per individual image by 'RGC Counter' or 'RGC Transduction', respectively. (A) Comparison between the number of RGCs counted by ImageJ plugins 'RGC Batch' and 'RGC Counter' from twelve images of injured and uninjured retinal wholemounts. [Y = 1x + 0, Pearson's r = 1, $R^2$ = 1, Pearson product-moment correlation]. (B) Comparison between the number of transduced RGCs measured by ImageJ plugins 'RGC Batch' and 'RGC Transduction' in twelve images from one transduced retinal wholemount. [Y = 1x + 0, Pearson's r = 1, $R^2$ = 1, Pearson product-moment correlation]. Each dot represents a separate image.

## Availability

### *Operating system*

Any operating system compatible with ImageJ or Fiji (http://imagej.net).

### *Programming language*

The 'Simple RGC' plugins are written in Kotlin (Version 1.3.50) and ImageJ is written in Java (Version 8 update 251 or later).

### *Additional system requirements*

No further requirements.

### *Dependencies*

The plugins are reliant on ImageJ (version 1.53b or later) and the following plugins: Auto Threshold (version 1.10.0 or later), fastcsv (version 1.0.3 or later), Apache POI (version 3.17 or later), Apache XmlBeans (version 3.1.0 or later), Bio Formats (Version 6.1.1. or later).

### *List of contributors*

Tiger Cross - Developed, tested and optimised plugins, wrote the manuscript.

Rasika Navarange - Developed, tested and optimised plugins.

Joon-Ho Son - Developed, tested and optimised plugins, wrote the manuscript.

William Burr - Developed, tested and optimised plugins.

Arjun Singh - Developed, tested and optimised plugins.

Kelvin Zhang - Developed and tested plugins.

Miruna Rusu - Tested plugins.

Konstantinos Gkoutzis – Reviewed plugin code.

Andrew Osborne - Performed biological experiments, wrote the manuscript, designed figure 1, supervised the study.

Bart Nieuwenhuis - Performed biological experiments, wrote the manuscript, designed figures 2-7 and supplementary figure 1, performed statistical analysis, supervised the study.

***Software location:***

 ***Code repository name:*** GitHub

 ***Identifier:*** https://github.com/sonjoonho/SimpleRGC

 ***Licence:*** GNU General Public Licence version 3

 ***Date published:*** 10/08/2020

 ***Language:*** English

**Reuse potential**

Although the 'Simple RGC' plugins have been optimised for inner retinal research, the plugins should also be able to quantify cell types other than RGCs within tissue sections. For instance, 'RGC counter' would be suitable for quantifying neuronal populations within the brain or dorsal root ganglia. Likewise, 'RGC Counter' could be used to quantify cells in vitro. 'RGC Transduction' could be re-used for non-RGC studies where a user may want to quantify the proportion of one particular cell type within a mixed cell population. The plugins should allow most cell types with a relatively round-like morphology to be investigated. Although this paper highlights usage of the plugins with mouse retinal tissue at 20x and 40x magnification, we have confirmed the plugins work over an extended range of microscope settings, and works on other species of animal. To increase the re-use potential, and allow other users to examine transduction in greater details, we have also enabled the plugin to measure fluorescence intensity of transduced cells in other channels. 'RGC Transduction' therefore can quantify expression of proteins tagged to a fluorescent marker which may be applicable to drug discovery platforms – measuring protein synthesis, activation, signalling or inhibition following treatment.

The authors of this manuscript are responsible for maintaining the plugins of 'Simple RGC'. We would encourage that potential bugs are reported by creating an issue on the GitHub account (by using the link: https://github.com/sonjoonho/SimpleRGC/issues), or alternatively by contacting the corresponding authors by e-mail.

**Funding statement**

This work was funded by the National Eye Research Centre (SAC 041), Fight For Sight UK, an ERA-NET NEURON grant AxonRepair by the Medical Research Council (MR/R004544/1) and the Department of Computing of Imperial College London via the Corporate Partnership Programme (CPP).

**Competing interests**

The authors declare that they have no competing interests.

## Supplementary figures

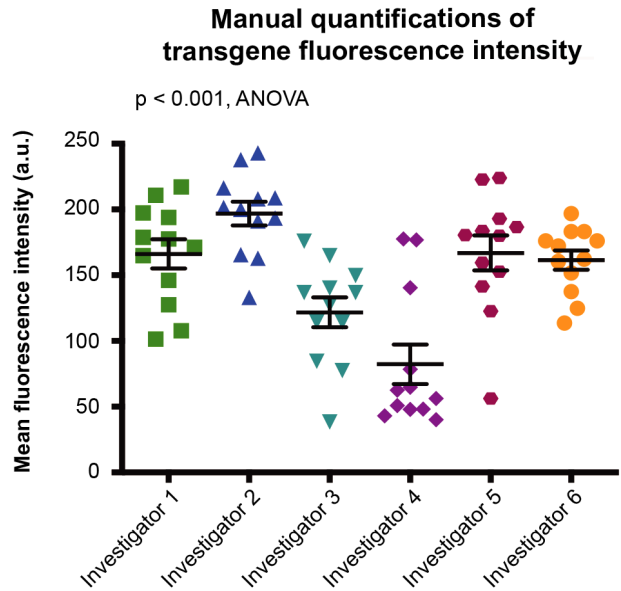

**Supplementary figure 1:** Variance in results of transgene fluorescence intensity when done by manual quantification. Six investigators independently analysed 12 fluorescence microscopy images originating from one transduced retinal wholemount. The mean fluorescence intensity per transduced RGC was measured by outlining 15 randomly selected RGCs per image in ImageJ. This method of manual quantification led to inconsistent outcomes between the investigators [$F_{(5, 66)} = 12.7$, $p < 0.001$, ANOVA], highlighting the need for automated quantification methods such as 'Simple RGC'.